# ACCELERATOR TECHNOLOGY FOR THE MANKIND

**Saleh Sultansoy**

*Gazi University, Ankara, Turkey and Institute of Physics, Baku, Azerbaijan*

**ABSTRACT**

Particle accelerators technology is one of the generic technologies which is locomotive of the development in almost all fields of science and technology. According to the U.S. Department of Energy: *Accelerators underpin every activity of the Office of Science and, increasingly, of the entire scientific enterprise. From biology to medicine, from materials to metallurgy, from elementary particles to the cosmos, accelerators provide the microscopic information that forms the basis for scientific understanding and applications. The combination of ground and satellite based observatories and particle accelerators will advance our understanding of our world, our galaxy, our universe, and ourselves.*

Because of this, accelerator technology should become widespread all over the world. Existing situation shows that a large portion of the world, namely the South and Mid-East, is poor on the accelerator technology. UNESCO has recognized this deficit and started SESAME project in Mid-East, namely Jordan. Turkic Accelerator Complex (TAC) project is more comprehensive and ambitious project, from the point of view of it includes light sources, particle physics experiments and proton and secondary beam applications.

At this stage, TAC project includes:

- Linac-ring type charm factory
- Synchrotron light source based on positron ring
- Free electron laser based on electron linac
- GeV scale proton accelerator
- TAC-Test Facility.

First part of this presentation is devoted to general status of particle accelerators around the world. The second part deal with the status of the TAC proposal.

## 1. INTRODUCTION

The title of this paper is inspired by the statement of the U.S. Department of Energy titled "Accelerator Technolgy for the Nation" [1]. Keeping in mind that particle accelerators are widely used in almost all fields of science and technology, accelerator technology should be included into R&D programme in all countries over the world. In this context, developed nations should support the knowledge transfer to developing ones. SESAME-like projects should be realized in all regions which are poor on the accelerator technology.

## 2. PARTICLE ACCELERATORS AROUND THE WORLD

The status of accelerator technology and its applications is reflected at large-scale biennial conferences: USA Particle Accelerator Conference (PAC), European Particle Accelerator Conference (EPAC) and Asian Particle Accelerator Conference (APAC). Proceedings of these conferences, as well as a number of other accelerator events, are collected at Joint Accelerator

Conferences Website [2]. An excellent review was presented at EPAC2000 by U. Amaldi [3] (see also [4]).

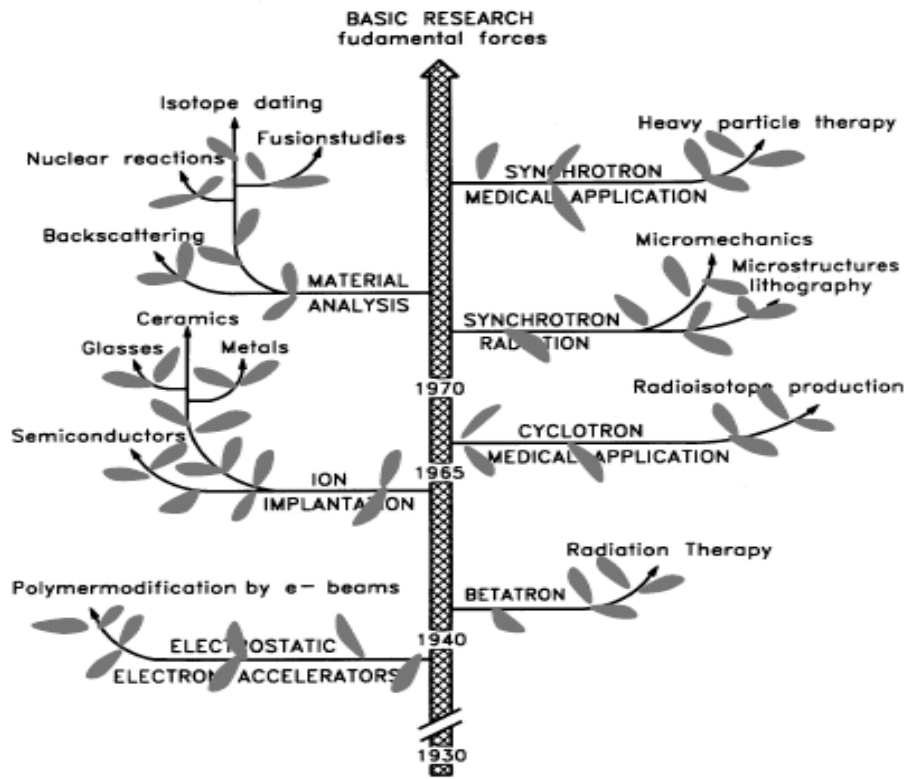

Figure 1: The *Time Tree* gives a pictorial view of the development of the applications of accelerators in both modification processes and sample analyses [3].

The classification of the 15000 accelerators around the world in 2000 according to application fields is given in the Table 1. In 1994 the total number of accelerators was about 10000, therefore, the progression rate is about 15% per year.

Table 1: Accelerators in the world [3].

| CATEGORY | NUMBER |
|---|---|
| Ion implanters and surface modifications | 7000 |
| Accelerators in industry | 1500 |
| Accelerators in non-nuclear research | 1000 |
| Radiotherapy | 5000 |
| Medical isotopes production | 200 |
| Hadrontherapy | 20 |
| Synchrotron radiation sorces | 70 |
| Research in nuclear and particle physics | 110 |
| **TOTAL** | **15000** |

## 2.1. FUNDAMENTAL RESEARCH: PARTICLE AND NUCLEAR PHYSICS

It is seen from the Table 1 that less than 1% of accelerators are used for fundamental research in particle and nuclear physics. The obvious flagman's are CERN [5], FNAL [6], DESY [7], SLAC [8] and KEK [9].

### 2.1.1. ENERGY FRONTIERS

An exploration of (multi-)TeV scale at constituent level is the main goal of High Energy Physics in a foreseen future. At the end of the last century, four ways to TeV scale, namely, ring type hadron machines, linear electron-positron machines, ring type muon colliders and linac-ring type lepton-hadron colliders were discussed (see [10] and references therein). Today, we deal with following situation:

- Hadron colliders. The LHC with 14 TeV center-of-mass energy will start hopefully in 2007 and the hundred-TeV energy VLHC is under consideration.
- Linear colliders. The CLIC is the sole machine with energy more than 1 TeV, and a 3 TeV center-of-mass energy is considered as third stage.
- Muon colliders. After the boom in 1990's, main activity is transferred to the $\nu$-factory options.
- Lepton-hadron colliders. The sole realistic way to (multi-)TeV scale is represented by linac-ring type machines.

Therefore, as the second way to (multi-)TeV scale, linac-ring type lepton-hadron colliders require more attention of the HEP community. Referring to reviews [10-14] for more details of these machines, as well as their additional $\gamma$p, eA, $\gamma$A and FEL$\gamma$A options, let me present here non-conventional approach to future energy frontiers for HEP: It may be well possible that, instead of constructing linear $e^+e^-$ colliders in the first stage, more attention must be paid to realizing linac-ring type ep colliders with the same electron beam energy (see Table 2).

Table 2: Energy Frontiers

| Colliders | Hadron | Lepton | Lepton-Hadron |
|---|---|---|---|
| **1990's** | Tevatron | SLC/LEP | HERA |
| √s, TeV | 2 | 0.1/0.1 →0.2 | 0.3 |
| L, $10^{31}cm^{-2}s^{-1}$ | 1 | 0.1/1 | 1 |
| **2010's** | LHC | "ILC"(TESLA) | "ILC"-LHC |
| √s, TeV | 14 | 0.5→1.0(0.8) | 3.7→5.3(4.7) |
| L, $10^{31}cm^{-2}s^{-1}$ | $10^3$ | $10^3$ | 1÷10 |
| **2020's** | VLHC | CLIC | "CLIC"-VLHC |
| √s, TeV | 200 | 3 | 34 |
| L, $10^{31}cm^{-2}s^{-1}$ | $10^3$ | $10^3$ | 10÷100 |

It is known that lepton-hadron collisions have been playing a crucial role in exploration of deep inside of matter. For example, the quark-parton model was originated from investigation of electron-nucleon scattering. The HERA with √s ≈ 0.3 TeV has opened a new era in this field extending the kinematics region by two orders both in high $Q^2$ and small x with respect to fixed target experiments. However, the region of sufficiently small x ($\leq 10^{-5}$) and simultaneously high $Q^2$ ($\geq 10$ GeV$^2$), where saturation of parton densities should manifest itself, is not currently achievable. The investigation of physics phenomena at extreme small x but sufficiently high $Q^2$ is very important for understanding the nature of strong interactions at all levels from nucleus to partons.

At the same time, the results from lepton-hadron colliders are necessary for adequate interpretation of physics at future hadron colliders. Concerning LHC, which hopefully will start in 2007, a √s ≈ 1 TeV ep collider will be very useful in earlier 2010's when precision era at LHC will begin.

Finally, multi-TeV center of mass energy ep colliders are competitive to future hadron and lepton colliders in search for the BSM physics.

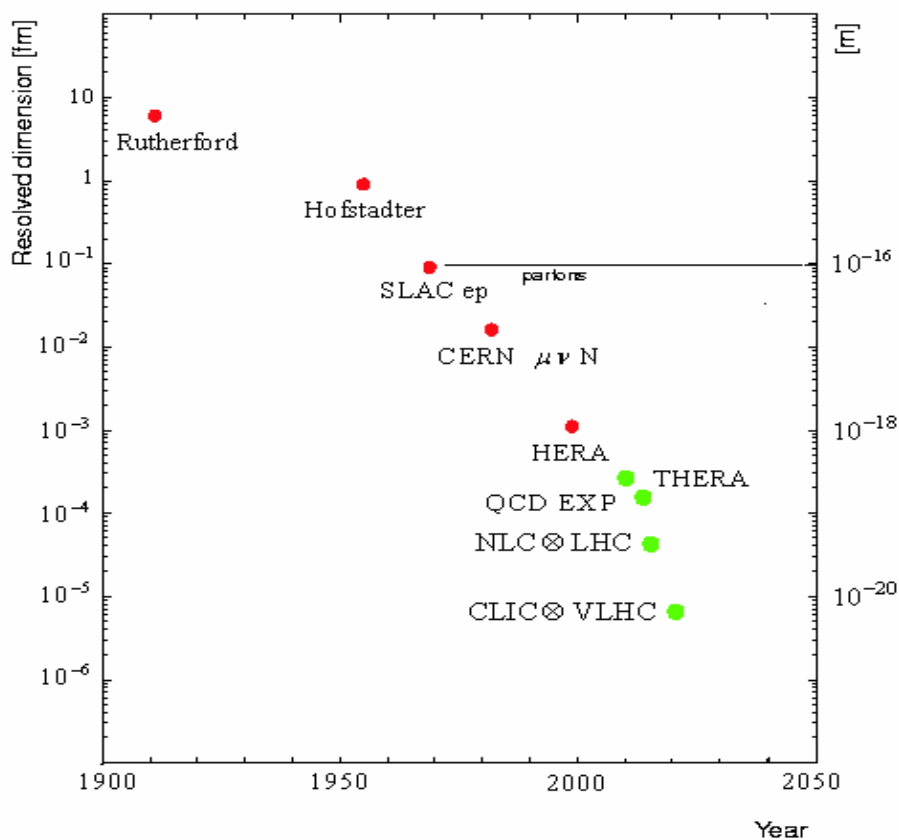

Figure 2: The development of the resolution power of the experiments exploring the inner structure of matter over time from Rutherford experiment to CLIC⊗VLHC.

### 2.1.2. PARTICLE FACTORIES

Particle factories represent rather high luminosity than high energy frontiers. The aim is copious production of known particles in order to investigate their properties in details. These machines can be grouped as electron-positron colliders (i.e. B, Charm, Tau factories) and secondary beams from high intensity proton accelerators (i.e. neutrino factories, meson factories etc). TAC Charm factory belongs to the first group.

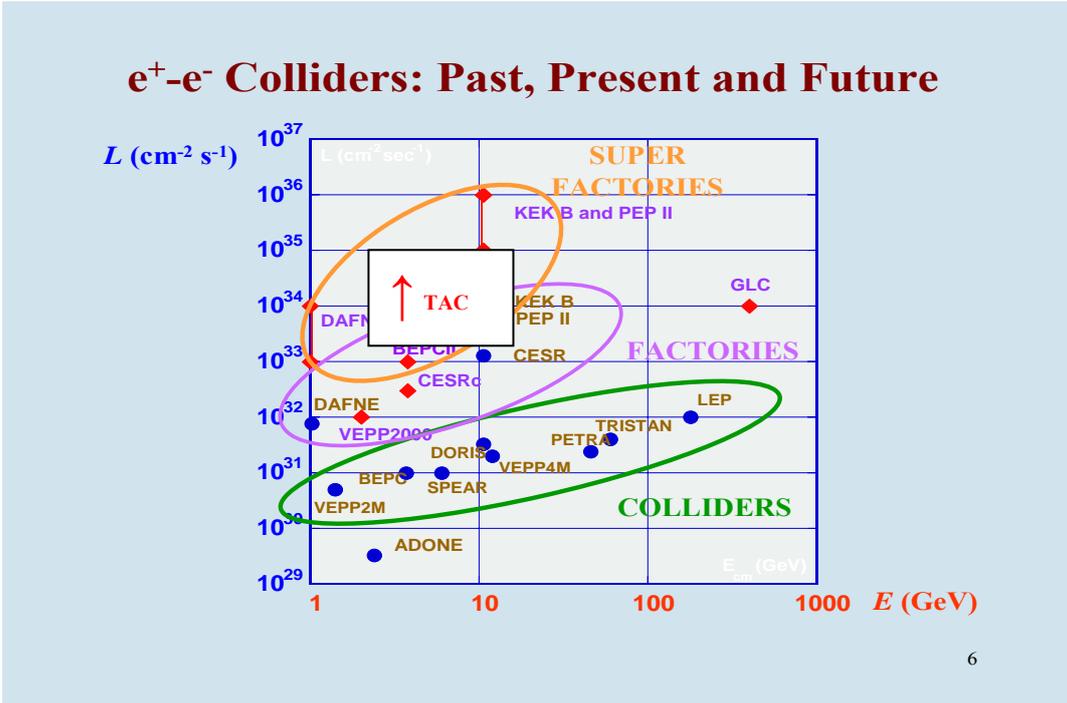

Figure 3: e⁺e⁻ particle factories [15]. The TAC Charm factory is added [16].

## 2.2. OTHER FIELDS OF SCIENCE AND TECHNOLOGY

It is seen from the Table 1 that more than 99% of accelerators are used in different fields of science and technology. Referring to reviews [3, 4] for more details let us mention that today two more rows should be added to the Table 1. Namely, neutron spallation sources and free electron lasers. The latter is called the fourth generation light source. Concerning the third generation light sources, namely synchrotron radiation, their map is presented in Figure 4.

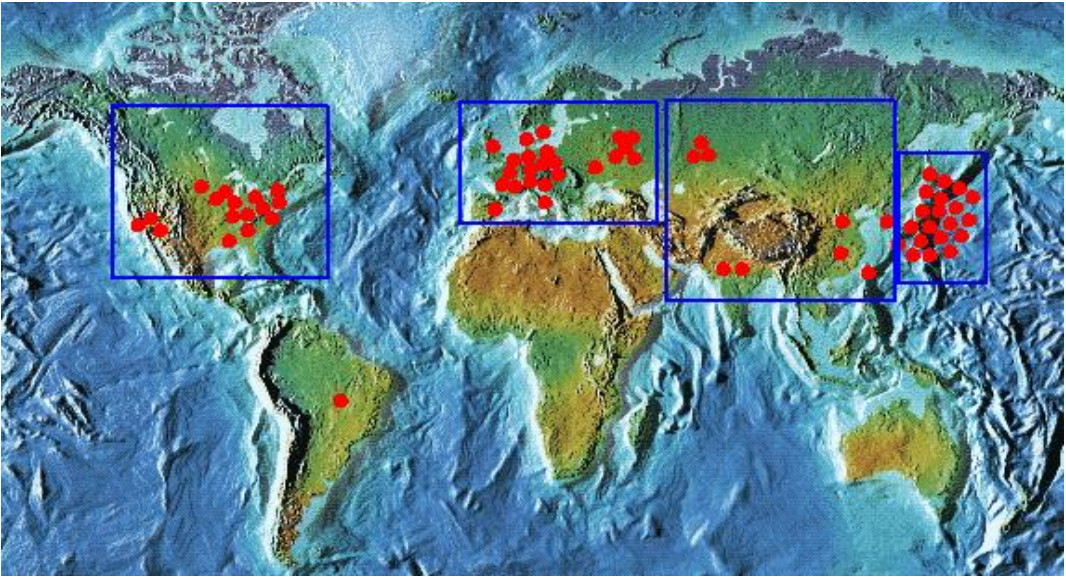

Figure 4: SR sources around the world.

# 3. THE STATUS OF TURKIC ACCELERATOR COMPLEX PROPOSAL

Approximately 10 years ago, linac-ring type charm-tau factory with synchrotron light source was proposed as a regional project for elementary particle physics [17]. Starting from 1997, a small group from Ankara and Gazi Universities begins a feasibility study for the possible accelerator complex in Turkey with the support of Turkish State Planning Organization (DPT) [18]. The results of the study is published in [19] and presented at EPACs [20, 21]. Starting from 2002, the conceptual design study of the TAC project has started with a relatively enlarged group (again with the DPT support). For the future plans, see the section on time schedule.

At this stage, TAC project includes:

- Linac-ring type charm factory
- Synchrotron light source based on positron ring
- Free electron laser based on electron linac
- GeV scale proton accelerator
- TAC-Test Facility.

The schematic view of the factory and light sources part of the Turkic Accelerator Complex is given in the Figure below.

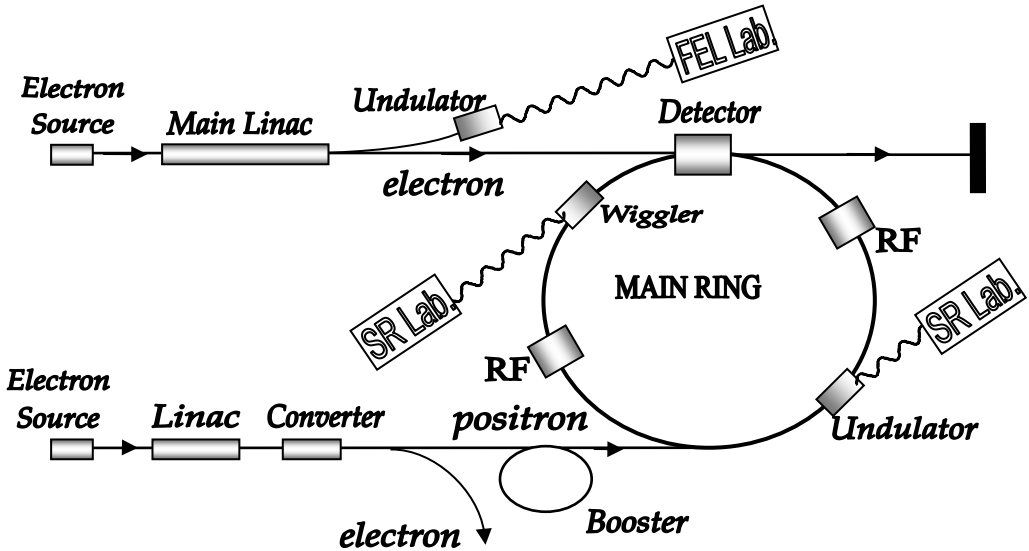

## 3.1. TAC CHARM FACTIRY

Up to now, we have analyzed linac-ring type $\phi$, charm and $\tau$ factory options. In principle L = $10^{34}$cm$^{-2}$s$^{-1}$ can be achieved for all three options. Concerning $\phi$ factory option, existing DA$\phi$NE $\phi$ factory has nominal L = $5 \times 10^{32}$cm$^{-2}$s$^{-1}$ and possible upgrades to higher luminosities are under consideration [22]. Therefore, physics search potential for the $\phi$ factory will be essentially exhausted before TAC commissioning. Concerning $\tau$ factory option, whereas $e^+e^- \to \tau^+\tau^-$ cross-section achieves to maximum value at $\sqrt{s}$ = 2.2 GeV, this advantage is dissipated with success of B-factories which has luminosity of $10^{34}$cm$^{-2}$s$^{-1}$ already. Moreover super B-factories with L = $10^{36}$cm$^{-2}$s$^{-1}$ are intensively discussed [23].

For these reasons, we inclined towards charm factory option. The center of mass energy is fixed by the mass of ψ (3770) resonance. Existing CLEO-c [24] works with $10^{32} cm^{-2} s^{-1}$. The BEP charm factory proposal [25] has design luminosity of $10^{33} cm^{-2} s^{-1}$. Therefore, TAC charm factory with $10^{34} cm^{-2} s^{-1}$, planned to work in mid 2010's, will contribute charm physics greatly. Differing from K and B mesons, where possible new physics manifest itself as a deviation from standard model (SM) background, D mesons has negligible SM background. The main parameter set for TAC charm factory is presented in Table 1. The restriction on luminosity coming from linac beam power can be relaxed by using of energy recovery linac. This topic is under study (recently the possibility of TAC Super-Carm Factory is discussed [16]).

Table 1: Tentative parameters of TAC charm factory

| Parameter | e⁻ linac | E⁺-ring |
|---|---|---|
| Energy, GeV | 1.00 | 3.56 |
| Particles per bunch, $10^{10}$ | 0.55 | 11.00 |
| β function at IP, cm | 0.45 | 0.45 |
| Normalized emittance, μm·rad | 6.17 | 22.00 |
| Bunch length, cm | 0.10 | 0.45 |
| Transverse size at IP, μm | 3.76 | 3.76 |
| Beam-beam tune shift | - | 0.056 |
| Collision frequency, MHz | 30 ||
| Luminosity ($H_D \cdot L$) | $1.4 \; 10^{34} \; cm^{-2} s^{-1}$ ||

The luminosity spectrum $dL/dW_{cm}$ obtained using GUINEA-PIG simulation program [29] with $\Delta E/E = 10^{-3}$ is plotted in Figure 2. It is seen that center of mass energy spread is well below $\Gamma_{\Psi(3S)} \approx 24$ MeV. Expected number of Ψ(3S) is about $10^9$ per working year ($10^7$ s). Let us remind that $D^+D^-$ and $D^0\bar{D}^0$ decay modes are dominant channels for Ψ(3S) decays.

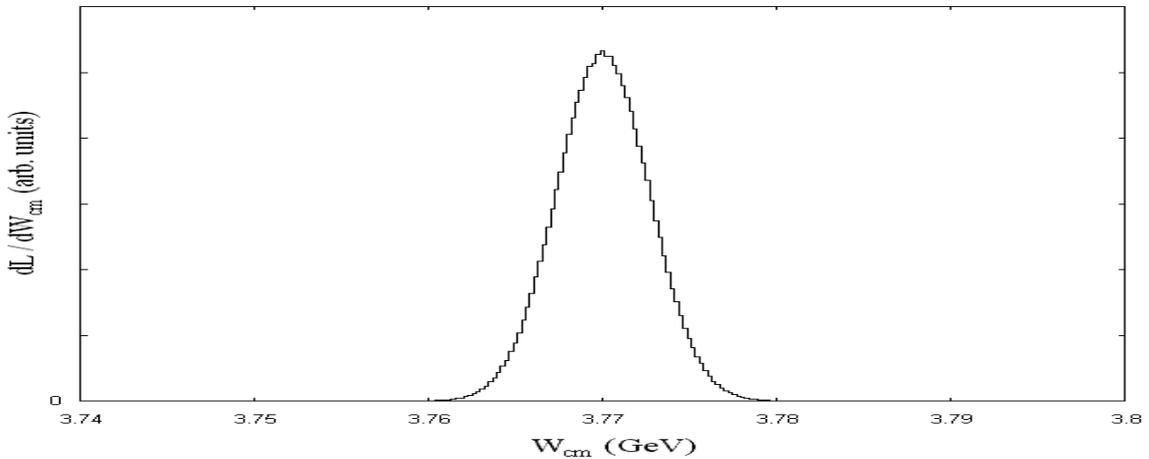

Figure 6: Luminosity spectrum for the TAC charm factory.

## 3.2. TAC SYNCHROTRON LIGHT SOURCE

Ref. [17] had considered additional positron storage ring dedicated to production of synchrotron radiation. Because of beam-beam tune shift restriction, the emittance of colliding beams in standard (ring-ring) type colliders inevitably should be chosen to be relatively large to obtain high luminosity:

$$L = f_c \frac{4\pi \gamma_p \gamma_e \Delta Q_p \Delta Q_e \varepsilon_p}{r_0^2 \beta_e^*} \qquad (1)$$

where subscript e (p) corresponds to electron (positron). This restricts the performance of synchrotron radiation obtained from insertion devices placed in standard type colliders.

Fortunately, this is not the case for linac-ring type machines. In this case, emittance of the positron beam does not essentially affect luminosity performance of the collider:

$$L = f_c \frac{\gamma_p \Delta Q_p N_p}{r_0 \beta_p^*} \qquad (2)$$

Therefore, the emittance of the positron beam can be chosen small enough to behave as a third generation light source in principle. Normalized emittance of the positron beam given in Table 1 corresponds to transverse emitance of 3 nm·rad, which is well below 20 nm·rad (upper limit for third generation SR sources).

Main parameters of TAC SR Facility were reported at EPAC 2000 [20]. Since then, construction of SESAME [26] has begun in Jordan and CANDLE [27] project has been developed in Armenia. For this reason, final decision on the number of insertion devices and beam lines of TAC SR Facility and their specifications will be made depending on realization of SESAME and CANDLE projects as well as on user potential in our region.

Several samples of optical beam lines design and related studies on TAC SR facility can be found in [28].

## 3.3. TAC FREE ELECTRON LASER

Main linac of the TAC charm factory can be operated separately to obtain FEL as seen from Figure 1. FEL operation is foreseen during the maintenance of the collider. With 1 GeV electron beam wave length of SASE FEL photons is expected to be 6.4 nm as in TTF-2 FEL. Detailed studies for different electron beam energies were presented at national conferences [28].

## 3.4. TAC PROTON ACCELERATOR

TAC proton accelerator proposal consists of 100÷300 MeV energy linear pre-accelerator and 1÷5 GeV main ring. The average beam current values for these machines would be ~30 mA and ~0.3 mA, respectively. Proton beams from two different points of the synchrotron will be forwarded to neutron and muon regions, where a wide spectrum of applied research is planned. In muon region, together with fundamental investigations such as test of QED and muonium-antimuonium oscillations, a lot of applied investigations such as High-T$_c$ superconductivity, phase transitions, impurities in semiconductors *et cetera* will be performed using the powerful Muon Spin Resonance (μSR) method. In neutron region investigations in different fields of applied physics, engineering, molecular biology and fundamental physics are planned.

## 3.5. TAC TEST FACILITY

Before building charm factory to obtain training of the young accelerator physicists and to get experience on accelerator technology on smaller scale, we plan to build infrared free electron laser (IR FEL) on 20÷50 MeV e-linac until 2009. IR FEL thought to work in oscillator mode. With undulator strength parameter K=1 and vertical distance between magnet poles g=3 cm, one can obtain wavelength values of 15 μm (IR FEL1) and 2 μm (IR FEL2) for 10 MeV and 50 MeV e-linac choices, respectively. Three experimental stations are planed to make research on biomedical subjects, semiconductor physics and photo chemical reactions.

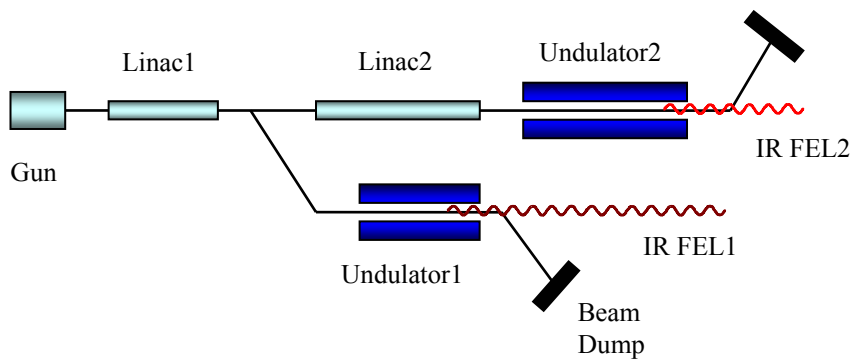

Figure 7: Schematic view of the TAC Test Facility.

## 3.5. TIME SCHEDULE

Tentative time schedule for the realization of TAC project follows:

2006:
- Completion of the conceptual design report
- Starting technical design study

2007:
- Construction of the building for TAC Test   Facility

2008:
- Installation of the TAC-TF linac

2009:
- Installation of the TAC-TF infra-red FEL and beam lines with the experimental stations
- Completion of the TAC technical design report

2010:
-  Commissioning of TAC-TF
-  Governmental decision on approval of TAC project

2015:
- Completion of factory and light source parts of TAC project.

2017:
- Completion of proton accelerator and experimental stations